*Article*

# Post-Quantum Key Agreement Protocols Based on Modified Matrix-Power Functions over Singular Random Integer Matrix Semirings

**Juan Pedro Hecht** [1], **Hugo Daniel Scolnik** [1,2,*]

[1] Master's degree in Information Security, Economic Science School, Exact and Natural Sciences School and Engineering School, University of Buenos Aires, phecht@dc.uba.ar

ORCID 0000-0002-4196-9059

[2] Computer Science Department, Faculty of Exact and Natural Sciences, University of Buenos Aires, hugo@dc.uba.ar

ORCID 0000-0002-0684-3661

**Abstract:** Post-quantum cryptography is essential for securing digital communications against threats posed by quantum computers. Researchers have focused on developing algorithms that can withstand attacks from both classical and quantum computers, thereby ensuring the security of data transmissions over public networks. A critical component of this security is the key agreement protocol, which allows two parties to establish a shared secret key over an insecure channel. This paper introduces two novel post-quantum key agreement protocols that can be easily implemented on standard computers using rectangular or rank-deficient matrices, exploiting the generalizations of the matrix power function, which is a generator of NP-hard problems. We provide basic concepts and proofs, pseudocodes, and examples, along with a discussion of complexity.

**Keywords:** key agreement protocol; non-commutative algebraic cryptography; post-quantum cryptography; rank- deficient matrices; matrix power function

## 1. Introduction

### *1.1 Post-Quantum Cryptography (PQC)*

Post-quantum cryptography focuses on developing algorithms that are resistant to quantum computer attacks [1]. Quantum machines can solve problems exponentially faster than classical computers [2], thereby compromising the current cryptographic methods. This urgency has led to the development of quantum-safe cryptographic algorithms for securing data transmission over public networks. A key application is the key agreement protocol [3] that enables two parties to establish a shared secret key over an insecure channel. This paper introduces post-quantum key agreement protocols using rectangular or rank-deficient matrices instead of square ones, leveraging generalizations of the matrix power function [4-9], which generates NP-hard problems [3]. These generalizations provide a novel approach for constructing secure key encapsulation mechanisms (KEMs) [10].

### 1.2 The history of post-quantum cryptography: Why and how?

Quantum computing originated in the early 1980s when Richard Feynman proposed using quantum mechanics to simulate physical processes [2]. The significance of quantum computing in cryptography was recognized in the mid-1990s, notably when Peter Shor introduced an algorithm in 1994 that factors large integers exponentially faster than any known classical algorithm [11]. This breakthrough has alarmed the cryptography community because many cryptographic protocols (e.g., RSA and ECC) depend on the difficulty of factoring large numbers or solving elliptic-curve discrete logarithmic problems [3]. Grover's quantum search algorithm further threatens symmetric encryption systems by accelerating brute-force attacks and reducing their security [12]. These developments highlight the potential threat that quantum computers pose to current cryptographic systems, thereby prompting the emergence of postquantum cryptography. Researchers are currently exploring mathematical problems that are resistant to both classical and quantum attacks, leading to lattice-based, code-based, multivariate, and hash-based cryptographic schemes [1]. These new protocols aim to secure communication against quantum threats while maintaining efficiency in classical hardware.

### 1.3 The Need for KEM Procedures in HTTPS

A critical component of contemporary Internet security is Hypertext Transfer Protocol Secure (HTTPS), which is employed to encrypt data exchanged between web browsers and servers. HTTPS utilizes the Transport Layer Security (TLS) protocol to secure connections, with handshaking functioning as the primary TLS process [13]. Handshaking establishes a shared secret key between the client (browser) and server, thereby facilitating secure communication over an insecure channel. Currently, handshaking relies on asymmetric cryptographic primitives such as RSA or elliptic curve cryptography (ECC) to exchange secret keys [3]. However, both the RSA and ECC are vulnerable to Shor's algorithm, indicating that a sufficiently powerful quantum computer can effectively compromise these cryptographic schemes, thereby jeopardizing the security of all HTTPS communications [1]. This presents a significant challenge, as HTTPS is extensively utilized for securing financial transactions, personal data, and other sensitive online information. Key Encapsulation Mechanisms (KEMs) have emerged as post-quantum alternatives to current key exchange mechanisms employed in HTTPS [10]. KEM facilitates the establishment of a shared secret key between two parties, even when communicating over an insecure network, by encapsulating the key within a ciphertext that can only be decrypted by the intended recipient. By implementing KEMs based on quantum-resistant problems, it is possible to secure the handshake process in TLS protocols, thereby ensuring that the security of HTTPS remains intact, even in the quantum-enabled future [1].

### *1.4 The Limitations of Quantum Key Distribution (QKD)*

Quantum Key Distribution (QKD) [2] is frequently regarded as a theoretically optimal solution to the key exchange problem; however, it has practical limitations that render it unsuitable for widespread implementation in standard client-server communications. QKD protocols such as BB84 and B92 [14] utilize quantum mechanical principles to provide an unbreakable key exchange, assuming the absence of implementation flaws. These protocols are demonstrably secure because any attempt to intercept the quantum key would perturb the quantum states, thereby alerting the communicating parties to the presence of an eavesdropper [2]. Notwithstanding its theoretical advantages, QKD is not feasible for securing common internet communication. QKD systems require specialized hardware, such as single-photon emitters and detectors, and typically rely on direct point-to-point optical fiber connections between communicating parties. For global, large-scale communication networks such as the Internet, where thousands or millions of clients connect to servers across vast distances, the requirement for dedicated quantum channels is both impractical and prohibitively expensive. Furthermore, the infrastructure required to support QKD is incompatible with current Internet infrastructure, which relies on a complex mesh of interconnected routers and switches. Although QKD is optimal for certain niche applications requiring the highest level of security (e.g., government or military communications), it is not a viable solution for securing everyday Internet traffic between clients and servers.

### 1.5 Post-Quantum KEM for Practical Quantum-Safe Security

Given the limitations of QKD, the development of post-quantum KEMs is a viable approach to securing Internet communication. In contrast to QKD, KEMs do not require specialized hardware and can be implemented using standard communication protocols. Through the design of KEMs that exploit the computational hardness of problems, such as those arising from matrix power functions with rectangular or rank-deficient matrices [15], it is possible to provide quantum-resistant security that integrates seamlessly into existing systems such as HTTPS and TLS.

### 1.6 NIST PQC Standardization Program

Recently, NIST (USA) concluded a PQC standardization program [16] by defining new standards for KEM exchanges and digital PQC signatures (FIPS 203, 204, 205) [17]; specifically, for key exchange, the selected protocols were Crystals-Kyber ML-KEM-512, ML-KEM-768, and ML-KEM-1024 [18]. However, this solution, although otherwise robust, is susceptible to AI-driven side-channel attacks [19]. Consequently, it is imperative to investigate alternative approaches, such as the algebraic-based solution presented herein, with particular emphasis on the latter. Furthermore, our protocols are significantly more code-compact than the learning-with-errors (LWE) lattices implemented by Crystals-Kyber, while maintaining the semantic security and quantum security levels mandated by NIST. Motivated by this objective, we developed novel KEMs that offer practical and resilient solutions for securing key exchanges in the post-quantum era. The proposed protocols can be implemented on a large scale, providing quantum-safe security for millions of users without necessitating substantial modifications to the internet infrastructure.

## 2. First protocol: using rectangular matrices

We define the Rectangular Matrix Power Function (RMPF), a generalization of the original Matrix Power Function (MPF) and outline its properties. We then present a Key Agreement Protocol (KAP) based on the RMPF [4-9] and, later, a second protocol, the Rank-Deficient Matrix Power Function (RDMPF) variant, which exploits rank-deficient matrices with standard linear algebra operations. We used the same notation as in the literature. Although we refer only to RMPF here, it should be noted that the same arguments apply to the RDMPF solution. Furthermore, recall that any rectangular matrix can be transformed into an equivalent square rank-deficient matrix by adding linear combinations of rows.

**Definition 1.** Equidimensional (m,n) rectangular matrices of integers (specifically p-prime $\mathbb{Z}p$ field elements) form an RM set, a (semi)ring structure with p-modular sums and p-modular Hadamard products (modular operations preserve numbers under a constant format).

**Definition 2.** Matrix elements of RM-set n powers are formed with the p-modular n powers of each element of the base matrix. Therefore, the product of the x-power of a W-element and y-power of the same element commutes (Wx.Wy = Wy.Wx) because the integer exponent products x and y commute. Hereafter, this paper deals only with RM sets when rectangular matrices are invoked.

**Definition 3.** Given any three matrices (X,W,C) of the same (m, n) RM set, the left-sided rectangular matrix power function (RMPF) exponential action of X over W is defined as the matrix C= $\{c_{ij}\}$:

$$\mathbf{X \rhd W \equiv {}^{X}W = C, \quad where \quad c_{ij} = \prod_{k=1}^{rank[X]} {w_{kj}}^{x_{ik}}} \qquad (1)$$

**Definition 4.** Given any three matrices (W, Y, D) of the same (m, n) RM set, the right RMPF exponential action of Y over W is equal to D; $\boldsymbol{D} = \{\boldsymbol{d_{ij}}\}$:

$$\mathbf{W} \lhd \mathbf{Y} \equiv \mathbf{W}^{\mathbf{Y}} = \mathbf{D}, \;\; \textbf{where} \;\; \mathbf{d_{ij}} = \prod_{\mathbf{l=1}}^{\mathbf{rank[Y]}} \mathbf{w_{il}}^{\mathbf{y_{lj}}} \tag{2}$$

**Definition 5.** Given any four matrices (X, W, Y, Q) of the same (m, n) RM set, the double-sided RPMF exponential action of matrix W with the left-sided X-matrix action and the right-sided Y-matrix action is defined as Q, such that $\boldsymbol{Q} = \{\boldsymbol{q_{ij}}\}$:

$$\boldsymbol{X} \rhd \boldsymbol{W} \lhd \boldsymbol{Y} \equiv {}^{\boldsymbol{X}}\boldsymbol{W}^{\boldsymbol{Y}} = \boldsymbol{Q}, \;\; \boldsymbol{where} \;\; \boldsymbol{q_{ij}} = \prod_{\boldsymbol{k=1}}^{\boldsymbol{rank(X)}} \prod_{\boldsymbol{l=1}}^{\boldsymbol{rank[Y]}} \boldsymbol{w_{kl}}^{\boldsymbol{x_{ik}} \cdot \boldsymbol{y_{lj}}} \tag{3}$$

**Lemma 1.** The RPMF is unilaterally associative, as Sakalauskas proved [7], if the following identities hold.

$$^{\mathbf{Y}}(^{\mathbf{X}}\mathbf{W}) = {}^{(\mathbf{YX})}\mathbf{W} = {}^{\mathbf{YX}}\mathbf{W} \;;\;\; (\mathbf{W}^{\mathbf{X}})^{\mathbf{Y}} = \mathbf{W}^{(\mathbf{XY})} = \mathbf{W}^{\mathbf{XY}} \tag{4}$$

and it is two-sided associative if:

$$(^{\mathbf{X}}\mathbf{W})^{\mathbf{Y}} = {}^{\mathbf{X}}(\mathbf{W}^{\mathbf{Y}}) = {}^{\mathbf{X}}\mathbf{W}^{\mathbf{Y}} \tag{5}$$

whereas the RMPF is defined as associative if both conditions hold.

**Lemma 2**. (m, n) RM sets obey the associative properties of an RMPF. This is a special case of Sakalauskas' proof [14] because the square (m, m) matrices are replaced by the particular case of (m,n) rectangular matrices.

**Lemma 3.** If (X, Y, U, V, W) are (m, n) RM-set matrices acting as one-sided (left or right) RMPF actions over another W and (X, U), (Y, V) are pairs of outer (ordinary) products, then both commutative conditions hold:

$$\mathbf{X}^{\mathbf{T}}.\mathbf{U} = \mathbf{U}^{\mathbf{T}}.\mathbf{X} \;\;;\;\; \mathbf{Y}^{\mathbf{T}}.\mathbf{V} = \mathbf{V}^{\mathbf{T}}.\mathbf{Y} \tag{6}$$

if the RMPFs of RM sets are associative (eq 4,5), then:

$$^{\mathbf{U}}(^{\mathbf{X}}\mathbf{W}^{\mathbf{Y}})^{\mathbf{V}} = {}^{\mathbf{UX}}\mathbf{W}^{\mathbf{YV}} = {}^{\mathbf{XU}}\mathbf{W}^{\mathbf{VY}} = {}^{\mathbf{X}}(^{\mathbf{U}}\mathbf{W}^{\mathbf{V}})^{\mathbf{Y}} \tag{7}$$

**Proof.** If pairwise outer products commute, the elements of their square product matrix exponents can be interchanged (see Definition 2, properties applied to Equations (1), (2), and (3)). Therefore, (7) holds.

**Lemma 4.** If $(\boldsymbol{\lambda 1}, \boldsymbol{\lambda 2}) \in \mathbb{Z}^{\mathbf{2}}$ $\boldsymbol{and}$ $(\boldsymbol{X}, \boldsymbol{U}, \boldsymbol{W})$ are members of the same RM set, then the scalar products $\boldsymbol{\lambda 1}.\boldsymbol{W} = \boldsymbol{X}$ $\boldsymbol{and}$ $\boldsymbol{\lambda 2}.\boldsymbol{W} = \boldsymbol{U}$ are matrices that satisfy condition (6).

**Proof.** Given an RM-set matrix $\boldsymbol{W} = \{\boldsymbol{w_{ij}}\}$, then $\boldsymbol{\lambda W} = \{\boldsymbol{\lambda w_{ij}}\}$ and $\boldsymbol{oW} = \{\boldsymbol{ow_{ij}}\}$, as $(\boldsymbol{\lambda}, \boldsymbol{o}) \in \mathbb{Z}^2$ then $\boldsymbol{\lambda w_{ij}}.\boldsymbol{ow_{ij}} = \boldsymbol{ow_{ij}}.\boldsymbol{\lambda w_{ij}} = (\boldsymbol{o}.\boldsymbol{\lambda})\boldsymbol{w_{ij}^2} = (\boldsymbol{\lambda}.\boldsymbol{o})\boldsymbol{w_{ij}^2}$ and therefore, condition (6, 7) holds.

**Lemma 5.** keyA = keyB

**Proof.** Considering (6) and (7), $\boldsymbol{KeyA} = \boldsymbol{A1} \rhd \boldsymbol{TB} \lhd \boldsymbol{B1} = \boldsymbol{A1} \rhd (\boldsymbol{A2} \rhd \boldsymbol{Base} \lhd \boldsymbol{B2}) \lhd \boldsymbol{B1} = \boldsymbol{A2} \rhd (\boldsymbol{A1} \rhd \boldsymbol{Base} \lhd \boldsymbol{B1}) \lhd \boldsymbol{B2} = \boldsymbol{A2} \rhd \boldsymbol{TA} \lhd \boldsymbol{B2} = \boldsymbol{KeyB}$.

## 3. Full description of the proposed RMPF action

*3.1 First schematic outline: Key-Agreement Protocol (KAP): see Figure 1.*

**Figure 1.** The proposed key agreement protocol, which is based on RMPF.

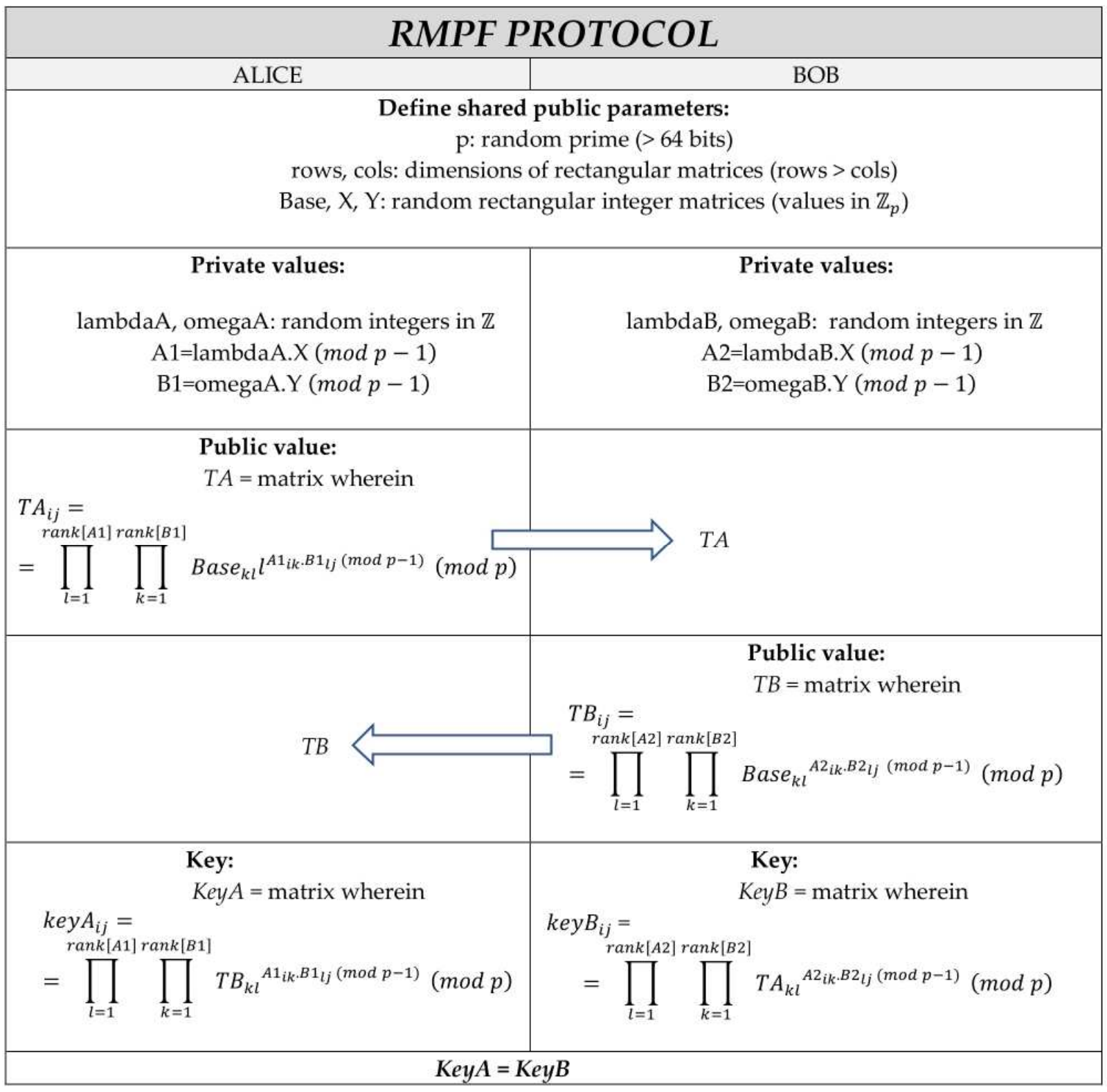

### *3.2 RMPF pseudocode description*

**Setup**

Both parties (Alice and Bob) agree :

A prime p.

The RM set dimensions (m, n), where m>n, and the applicable operations are governed by Hadamard algebra.

The parties share three RM random matrices with bases, X and Y, with values in ℤp.

**Alice's private values**

lambdaA, omegaA: random numbers in ℤ

A1=lambdaA.X (mod p-1), B1=omegaA.Y (mod p-1)

**Alice's public token**

Generate the TA matrix; {TAij }

$$\mathrm{TA_{ij}} = \prod_{l=1}^{\mathrm{rank[A1]}} \prod_{k=1}^{\mathrm{rank[B1]}} Base_{kl}^{\mathrm{A1_{ik} \cdot B1_{lj}}\ (mod\ p-1)}\ (mod\ p)$$

TA is sent to Bob.

**Bob's private values**

lambdaB, omegaB: random numbers in ℤ

A2=lambdaB.X (mod p-1), B2=omegaB.Y (mod p-1)

**Bob's public token**

Generate the TB matrix; {TBij }

$$TB_{ij} = \prod_{l=1}^{rank[A2]} \prod_{k=1}^{rank[B2]} Base_{kl}^{A2_{ik} \cdot B2_{lj} \,(mod\, p-1)} \;(mod\, p)$$

TB is sent to Alice.

**Shared key**

Alice and Bob generate keyA and keyB matching matrices, respectively. {keyAij, keyBij}

$$keyA_{ij} = \prod_{l=1}^{rank[A1]} \prod_{k=1}^{rank[B1]} TB_{kl}^{A1_{ik} \cdot B1_{lj} \,(mod\, p-1)} \;(mod\, p)$$

$$keyB_{ij} = \prod_{l=1}^{rank[A2]} \prod_{k=1}^{rank[B2]} TA_{kl}^{A2_{ik} \cdot B2_{lj} \,(mod\, p-1)} \;(mod\, p)$$

*3.3 A toy numeric example of the RMPF Key-Agreement Protocol (KAP)*

This example follows the first protocol.

Defining prime as $\boldsymbol{p = 65537}$, $\mathbf{rows = 5}$, $\mathbf{cols = 3}$, it follows:

$$\mathbf{Base} = \begin{pmatrix} 44664 & 10605 & 58177 \\ 37079 & 44866 & 49280 \\ 45409 & 15609 & 726 \\ 57731 & 9471 & 41234 \\ 52116 & 32253 & 872 \end{pmatrix} \quad (8)$$

$$\mathbf{X} = \begin{pmatrix} 25454 & 62439 & 63614 \\ 39060 & 9694 & 46468 \\ 6392 & 43055 & 15148 \\ 26377 & 42964 & 30474 \\ 55812 & 12484 & 59987 \end{pmatrix} \quad (9)$$

$$\mathbf{Y} = \begin{pmatrix} 32239 & 25090 & 32249 \\ 31554 & 15896 & 40908 \\ 53836 & 29341 & 55133 \\ 49046 & 44776 & 7840 \\ 53994 & 48994 & 62776 \end{pmatrix} \quad (10)$$

$$\mathbf{lambdaA} = \mathbf{60308} \quad (11)$$

$$\mathbf{A1} = \begin{pmatrix} 30104 & 3724 & 21208 \\ 4496 & 44632 & 7248 \\ 5984 & 24620 & 39280 \\ 54324 & 41616 & 65480 \\ 46672 & 7504 & 43260 \end{pmatrix} \quad (12)$$

$$\mathbf{omegaA} = \mathbf{36605} \quad (13)$$

$$\mathbf{B1} = \begin{pmatrix} 1843 & 63482 & 40213 \\ 27706 & 44472 & 5276 \\ 64796 & 23337 & 27881 \\ 35646 & 35656 & 1056 \\ 15682 & 32730 & 26712 \end{pmatrix} \quad (14)$$

$$\mathbf{TokenA} = \begin{pmatrix} 19050 & 55225 & 32116 \\ 20307 & 33635 & 46068 \\ 50694 & 64046 & 51330 \\ 1754 & 3460 & 4352 \\ 50272 & 26460 & 52031 \end{pmatrix} \quad (15)$$

$$\mathbf{lambdaB} = \mathbf{25401} \quad (16)$$

$$\mathbf{A2} = \begin{pmatrix} 44414 & 41839 & 3598 \\ 13556 & 18542 & 30308 \\ 30520 & 40823 & 12492 \\ 27649 & 23092 & 24378 \\ 5860 & 42916 & 17787 \end{pmatrix} \quad (17)$$

$$\mathbf{omegaB} = \mathbf{64763} \quad (18)$$

$$\mathbf{B2} = \begin{pmatrix} 48469 & 4086 & 40739 \\ 53686 & 33160 & 32004 \\ 132 & 60399 & 46127 \\ 32786 & 56696 & 34528 \\ 9070 & 7446 & 36328 \end{pmatrix} \quad (19)$$

$$\mathbf{TokenB} = \begin{pmatrix} 8616 & 10721 & 1187 \\ 43735 & 40329 & 8281 \\ 52007 & 53646 & 42109 \\ 20747 & 37614 & 61557 \\ 18153 & 19017 & 15289 \end{pmatrix} \quad (20)$$

$$\mathbf{KeyA} = \begin{pmatrix} 23030 & 13518 & 44672 \\ 8819 & 10151 & 12163 \\ 21 & 40471 & 6436 \\ 45352 & 62662 & 60452 \\ 9532 & 30007 & 11905 \end{pmatrix} \quad (21)$$

$$\mathbf{KeyB} = \begin{pmatrix} 23030 & 13518 & 44672 \\ 8819 & 10151 & 12163 \\ 21 & 40471 & 6436 \\ 45352 & 62662 & 60452 \\ 9532 & 30007 & 11905 \end{pmatrix} \quad (22)$$

*3.4. Real life parameters.*

The success rate of a brute-force attack exponentially decreases as the matrix order increases. In this context, however, this phenomenon is not applicable because the (X, Y) matrices are public, and security relies on secret lambda and omega integers. It is recommended to utilize RMPF Protocol matrices with a rank on the order of 100 and p ~ 2^64 as the minimum values. Consequently, any attack against two random integers in $\mathbb{Z}$p corresponds to a 128-bit brute force attack.

## 4. Second Protocol using rank-deficient matrices

*4.1. Second schematic outline: Multi-round Key-Agreement Protocol (KAP): see Figure 2.*

**Figure 2.** The multi-round Key-Agreement Protocol, which is based on the RDMPF.

| ***Multi-RDMPF PROTOCOL*** | |
|---|---|
| ALICE | BOB |
| **Define shared public parameters:**<br>p: random prime<br>dim: dimension of all square matrices (dim = rows = cols)<br>W: full-rank integer matrix (values in $\mathbb{Z}_p$)<br>BaseXU, BaseYV: rank-deficient matrices (values in $\mathbb{Z}_p$)<br>expMax: top exponent value $\epsilon$ Z<br>*rounds*: repetitions of key generation sessions $\epsilon$ Z<br>Alist=Blist=KeyAlist=KeyBlist={} (auxiliary values) | |
| **Repeat** *rounds* **times** | |
| **Private values:**<br>randX, randY: random integers in $\mathbb{Z}_{expMax}$<br>X=BaseXU$^{\text{randX}}$ (mod *p-1*),<br>Y=BaseYV$^{\text{randY}}$ (mod *p-1*), | **Private values:**<br>randU, randV: random integers in $\mathbb{Z}_{expMax}$<br>U=BaseXU$^{\text{randU}}$ (mod *p-1*),<br>V=BaseYV$^{\text{randV}}$ (mod *p-1*), |
| **Public value:**<br>*TA* = matrix wherein<br>$TA_{ij} = \\ = \prod_{l=1}^{dim}\prod_{k=1}^{dim} W_{kl}{}^{X_{ik}\cdot Y_{lj}\ (mod\ p-1)}\ (mod\ p)$<br>*Append TA to Alist until* **End Repeat** *and send to Bob* | *Alist* |
| *Blist* | **Public value:**<br>*TB* = matrix wherein<br>$TB_{ij} = \\ = \prod_{l=1}^{dim}\prod_{k=1}^{dim} W_{kl}{}^{U_{ik}\cdot V_{lj}\ (mod\ p-1)}\ (mod\ p)$<br>*Append TB to Blist until* **End Repeat** *and send to Alice* |
| **round-Key** (obtained from round-TB, **parsed** round-by-round from Blist)<br>*KeyA* = matrix wherein<br>$keyA_{ij} = \\ = \prod_{l=1}^{dim}\prod_{k=1}^{dim} TB_{kl}{}^{X_{ik}\cdot Y_{lj}\ (mod\ p-1)}\ (mod\ p)$<br>*Append keyA to Alist (rounds times)* | **round-Key** (obtained from round-TA, **parsed** round-by-round from Alist)<br>*KeyB* = matrix wherein<br>$keyB_{ij} = \\ = \prod_{l=1}^{dim}\prod_{k=1}^{dim} TA_{kl}{}^{U_{ik}\cdot V_{lj}\ (mod\ p-1)}\ (mod\ p)$<br>*Append keyB to Bist (rounds times)* |
| ***Session Keys (512-bits)***<br>*Alist=Blist*<br>***ALICE Key = SHA3-512(Alist)***<br>***BOB Key = SHA3-512(Blist)*** | |

### *4.2. RDMPF pseudocode description*

We developed the multi-PQC-RDMPF-KAP algorithm, which generally obeys our rectangular protocol with the following variants:

- o Replace rectangular matrices with equivalent rank-deficient matrices.
- o Replace the use of Hadamard algebra with conventional linear algebra, which reduces the feasibility of linearization attacks by replacing numerical powers in private matrices with multilinear inner products.

o Operate in rounds and consolidate partial keys into a combined key.

o Add a cryptographically secure hash function (SHA3-512) to create a 512-bit shared session key, thereby adding a random oracle to the protocol [21].

Here we describe one round of the rank-deficient algorithm:

**Public setup parameters:**

p prime

dim=rows=cols $\epsilon$ Z

Full-Rank nucleus matrix: W (mod p)

Rank-deficient matrices {BaseXU, BaseYV } (mod p)

Rounds $\epsilon$ Z

expMax $\epsilon$ Z

**Alice's private keys:**

$X=BaseXU^{randX}$ (mod p-1), $Y=BaseYV^{randY}$ (mod p-1)

where {randX, randY} $\epsilon$ $Z_{expMax}$

**Alice's public key: (token for Bob):**

TA=RDMPF(X,W,Y) (mod p) ; RDMPF( ) is defined below.

**Bob's private keys: :**

$U=BaseXU^{randU}$ (mod p-1), $V=BaseYV^{randV}$ (mod p-1)

where {randU, randV} $\epsilon$ $Z_{expMax}$

**Bob's public key: (token for Alice):**

TB=RDMPF(U,W,V) (mod p) ; RDMPF( ) is defined below.

*Remark 1: the sporadic occurrence of null matrices at the token level should be controlled, in which case the round is restarted. Once the tokens have been exchanged, both can compute the same key, Alice's key is KeyA=RDMPF(X,TB,Y) (mod p), and Bob's Key is KeyB=RDMPF(U,TA,V) (mod p)*

**The tokens (public keys) of the rounds are concatenated sequentially, row by row, and exchanged:**

Alice: Alist= A1...Arounds, and Bob: Blist= B1...Brounds

Based on the tokens (public keys) received from the counterpart and retrieved round-by-round, the respective keys of each round are obtained (see *Remark 1).*

**The keys are concatenated sequentially row by row:**

Alice: KeyAlist= KeyA1...KeyArounds, and Bob: Blist= KeyB1...KeyBrounds

**The following 512-bit session keys were obtained:**

KsessionA = SHA3-512(KeyAlist)

KsessionB = SHA3-512(KeyBlist)

Here we describe the Rank Deficient Matrix Power Function (RDMPF) using Mathematica (pseudocode-like) format. This algorithm illustrates the RDMPF in full detail. Note that there is no simple description of the entire matrix, and only element-by-element definitions. This function can be interpreted as the inner product exponential version of the ordinary matrix product that occurs when left and right matrix actions are applied. Expressed in simple terms, ordinary sums are transformed into products and ordinary products are transformed into exponential operations. This is the source of the overall complexity, which blocks both classical and quantum attacks.

```
1.  RDMPFint[X_,W_,Y_]:= Module[{i, j, L, K, z, ex, pr},   (* local variables *)
2.  Q=Rmat[rows, cols, prime];   (* modular random integer matrix, here rows=cols=dim *)
3.  Do[
4.     Do[
5.        pr=1;
6.        Do[
7.           Do[
8.              ex=Mod[Times[X[[i,K]],Y[[L,j]]], prime-1];   (*modular product *)
9.              z=IntFastPower[W[[K,L]], ex];   (* square-and-multiply power function*)
10.             pr=Times[pr, z];
11.          {L, 1, cols}];   (* end Do *)
12.       {K, 1, cols}];   (* end Do *)
13.       Q[[i, j]]=Mod[pr, prime];
14.    {j, cols}]   (* end Do *)
15. {i, rows}];   (* end Do *)
16. Q]; (* output *)
```

A powerful variant of the RDMPFint[X, W, Y] function to be implemented in real-life applications is to incorporate in the exponents the product with an arbitrary random 'sigma' integer, which does not modify the correct session keys generation. Therefore, we suggest replacing line 8 of the algorithm with the following:

```
8.              ex=Mod[Times[sigma, Times[X[[i,K]],Y[[L,j]]]], prime-1];
```

and sigma must be defined as a constant value at the start of the session as sigma=RandomInteger[{-anyLimit, +anyLimit}]. Multiple advantages were obtained:

a) Stronger randomized version than the original MPF.
b) Block linearization of the function through a logarithmic version that transforms power into products and products into sums.
c) Create a potential subliminal channel between the parts.

Here is a small symbolic example:

1. dim=rows=cols=2; $\sigma$=random integer
2. W={{$w_{11}$,$w_{12}$},{$w_{21}$,$w_{22}$}};
3. X={{$x_{11}$, $x_{12}$},{$x_{21}$,$x_{22}$}};
4. Y={{$y_{11}$,$y_{12}$},{$y_{21}$,$y_{22}$}};

5. **Q=RDMPF[X,W,Y]**;

6. Q =
7. {{$w_{11}^{\sigma x11\, y11}\, w_{12}^{\sigma x11\, y21}\, w_{21}^{\sigma x12\, y11}\, w_{22}^{\sigma x12\, y21}$, $w_{11}^{\sigma x11\, y12}\, w_{12}^{\sigma x11\, y22}\, w_{21}^{\sigma x12\, y12}\, w_{22}^{\sigma x12\, y22}$},
8. {$w_{11}^{\sigma x21\, y11}\, w_{12}^{\sigma x21\, y21}\, w_{21}^{\sigma x22\, y11}\, w_{22}^{\sigma x22\, y21}$, $w_{11}^{\sigma x21\, y12}\, w_{12}^{\sigma x21\, y22}\, w_{21}^{\sigma x22\, y12}\, w_{22}^{\sigma x22\, y22}$}}

### *4.3 A toy numeric example of the multi RDMPF Key-Agreement Protocol (KAP)*

This example follows the second protocol. Rows repetitions (here 1, 2) are at random positions.

Defining prime as $\boldsymbol{p} = \mathbf{65537}$, $\mathbf{dim} = \mathbf{5}$, **expMax=10000, rounds=2**, it follows:

$$\boldsymbol{W} = \begin{pmatrix} 36671 & 1524 & 19050 & 12061 & 61140 \\ 5366 & 34773 & 37275 & 10709 & 60768 \\ 59994 & 8372 & 16513 & 19213 & 18024 \\ 22554 & 1387 & 10646 & 57542 & 54414 \\ 62130 & 15684 & 5868 & 17933 & 2855 \end{pmatrix} \qquad (23)$$

$$\boldsymbol{BaseXU} = \begin{pmatrix} 57543 & 23480 & 42992 & 19549 & 59890 \\ 57543 & 23480 & 42992 & 19549 & 59890 \\ 43343 & 28960 & 64751 & 37741 & 48337 \\ 1091 & 62357 & 30242 & 50955 & 3101 \\ 37839 & 36136 & 38757 & 10107 & 12470 \end{pmatrix} \qquad (24)$$

$$\mathbf{BaseYV} = \begin{pmatrix} 61098 & 25692 & 18347 & 31256 & 2737 \\ 61098 & 25692 & 18347 & 31256 & 2737 \\ 23628 & 47854 & 30452 & 10898 & 3201 \\ 4055 & 43906 & 6517 & 25648 & 29018 \\ 13622 & 59502 & 23730 & 40601 & 41483 \end{pmatrix} \quad (25)$$

First round

$$\mathbf{randX} = 4267 \quad (26)$$

$$\mathbf{X} = \begin{pmatrix} 25300 & 53591 & 3358 & 6302 & 15971 \\ 25300 & 53591 & 3358 & 6302 & 15971 \\ 59640 & 62777 & 43012 & 50996 & 33510 \\ 8272 & 23015 & 13985 & 6756 & 47019 \\ 64853 & 6353 & 9303 & 16909 & 11272 \end{pmatrix} \quad (27)$$

$$\mathbf{randY} = 4651 \quad (28)$$

$$\mathbf{Y} = \begin{pmatrix} 50294 & 15396 & 2447 & 20604 & 46313 \\ 50294 & 15396 & 2447 & 20604 & 46313 \\ 52856 & 57814 & 29792 & 40618 & 1969 \\ 25287 & 53714 & 4577 & 4384 & 26014 \\ 24014 & 4806 & 32294 & 48601 & 23187 \end{pmatrix} \quad (29)$$

$$\mathbf{randU} = 6066 \quad (30)$$

$$\mathbf{U} = \begin{pmatrix} 61917 & 24420 & 29078 & 47059 & 18070 \\ 61917 & 24420 & 29078 & 47059 & 18070 \\ 35935 & 20952 & 51333 & 41093 & 16163 \\ 41155 & 1979 & 10882 & 17171 & 37033 \\ 38861 & 15750 & 29077 & 7509 & 61114 \end{pmatrix} \quad (31)$$

$$\mathbf{randV} = 8472 \quad (32)$$

$$\mathbf{V} = \begin{pmatrix} 37353 & 1020 & 59757 & 44920 & 18981 \\ 37353 & 1020 & 59757 & 44920 & 18981 \\ 38256 & 24936 & 25399 & 44464 & 10051 \\ 36307 & 16166 & 52913 & 49849 & 13652 \\ 51670 & 11528 & 54954 & 50615 & 6058 \end{pmatrix} \quad (33)$$

$$\mathbf{Token\ A} = \begin{pmatrix} 53838 & 27572 & 60974 & 49207 & 54423 \\ 53838 & 27572 & 60974 & 49207 & 54423 \\ 7986 & 15752 & 8069 & 40416 & 15771 \\ 41410 & 8254 & 42646 & 57132 & 64087 \\ 62119 & 17840 & 19622 & 20589 & 6234 \end{pmatrix} \quad (34)$$

$$\mathbf{Token\ B} = \begin{pmatrix} 29348 & 1649 & 29136 & 53009 & 60590 \\ 29348 & 1649 & 29136 & 53009 & 60590 \\ 47894 & 18698 & 17669 & 19542 & 31170 \\ 5356 & 9122 & 3736 & 17535 & 33881 \\ 46266 & 10907 & 21467 & 39257 & 36010 \end{pmatrix} \quad (35)$$

$$\mathbf{KeyA} = \begin{pmatrix} 20743 & 10836 & 64775 & 35222 & 44472 \\ 20743 & 10836 & 64775 & 35222 & 44472 \\ 49310 & 2062 & 65040 & 46960 & 24883 \\ 40381 & 25492 & 38040 & 58289 & 65195 \\ 43404 & 25602 & 54209 & 59994 & 36225 \end{pmatrix} \quad (36)$$

$$\textbf{KeyB} = \begin{pmatrix} 20743 & 10836 & 64775 & 35222 & 44472 \\ 20743 & 10836 & 64775 & 35222 & 44472 \\ 49310 & 2062 & 65040 & 46960 & 24883 \\ 40381 & 25492 & 38040 & 58289 & 65195 \\ 43404 & 25602 & 54209 & 59994 & 36225 \end{pmatrix} \quad (37)$$

<u>Second round</u>

$$\textbf{randX} = \textbf{6171} \quad (38)$$

$$\textbf{X} = \begin{pmatrix} 20687 & 43044 & 29876 & 65277 & 34570 \\ 20687 & 43044 & 29876 & 65277 & 34570 \\ 48043 & 42272 & 30547 & 16281 & 53097 \\ 64011 & 43209 & 15826 & 58203 & 65225 \\ 59031 & 50156 & 13641 & 54627 & 6418 \end{pmatrix} \quad (39)$$

$$\textbf{randY} = \textbf{2414} \quad (40)$$

$$\textbf{Y} = \begin{pmatrix} 40891 & 39362 & 36749 & 34923 & 28810 \\ 40891 & 39362 & 36749 & 34923 & 28810 \\ 22763 & 63190 & 28195 & 33540 & 27134 \\ 56708 & 35280 & 14969 & 48184 & 42201 \\ 38364 & 57222 & 24807 & 17310 & 52808 \end{pmatrix} \quad (41)$$

$$\textbf{randU} = \textbf{7574} \quad (42)$$

$$\textbf{U} = \begin{pmatrix} 61547 & 33968 & 56069 & 41953 & 50743 \\ 61547 & 33968 & 56069 & 41953 & 50743 \\ 29714 & 32573 & 36652 & 42508 & 7927 \\ 33931 & 35041 & 24823 & 50021 & 61711 \\ 38392 & 28428 & 60085 & 13340 & 4043 \end{pmatrix} \quad (43)$$

$$\textbf{randV} = \textbf{1456} \quad (44)$$

$$\textbf{V} = \begin{pmatrix} 17998 & 3012 & 8841 & 26426 & 43907 \\ 17998 & 3012 & 8841 & 26426 & 43907 \\ 49148 & 8686 & 26452 & 55316 & 51969 \\ 64313 & 53978 & 52641 & 4196 & 14662 \\ 51704 & 8754 & 12104 & 61813 & 36643 \end{pmatrix} \quad (45)$$

$$\textbf{Token A} = \begin{pmatrix} 21108 & 54710 & 20029 & 6255 & 14963 \\ 21108 & 54710 & 20029 & 6255 & 14963 \\ 28723 & 28942 & 9398 & 51028 & 3356 \\ 44003 & 6940 & 4827 & 50400 & 35084 \\ 54653 & 19386 & 46270 & 24516 & 19667 \end{pmatrix} \quad (46)$$

$$\textbf{Token B} = \begin{pmatrix} 31055 & 8992 & 38240 & 47046 & 52571 \\ 31055 & 8992 & 38240 & 47046 & 52571 \\ 53708 & 5236 & 39748 & 56283 & 63932 \\ 27273 & 31500 & 58981 & 63915 & 16157 \\ 21773 & 26963 & 14715 & 52520 & 13589 \end{pmatrix} \quad (47)$$

$$\textbf{KeyA} = \begin{pmatrix} 33253 & 42632 & 21998 & 52285 & 49951 \\ 33253 & 42632 & 21998 & 52285 & 49951 \\ 14086 & 35325 & 53116 & 60717 & 41037 \\ 3238 & 39606 & 1643 & 48792 & 26310 \\ 19481 & 30394 & 40594 & 46821 & 12282 \end{pmatrix} \quad (48)$$

$$\textbf{KeyB} = \begin{pmatrix} \mathbf{33253} & \mathbf{42632} & \mathbf{21998} & \mathbf{52285} & \mathbf{49951} \\ \mathbf{33253} & \mathbf{42632} & \mathbf{21998} & \mathbf{52285} & \mathbf{49951} \\ \mathbf{14086} & \mathbf{35325} & \mathbf{53116} & \mathbf{60717} & \mathbf{41037} \\ \mathbf{3238} & \mathbf{39606} & \mathbf{1643} & \mathbf{48792} & \mathbf{26310} \\ \mathbf{19481} & \mathbf{30394} & \mathbf{40594} & \mathbf{46821} & \mathbf{12282} \end{pmatrix} \quad (49)$$

$$\boldsymbol{A - Token\ combined\ list} = \{\mathbf{53838}, \dots (\mathbf{48}\ \boldsymbol{terms}) \dots, \mathbf{19667}\} \quad (50)$$

$$\boldsymbol{B - Token\ combined\ list} = \{\mathbf{29348}, \dots (\mathbf{48}\ \boldsymbol{terms}) \dots, \mathbf{13589}\} \quad (51)$$

$$\boldsymbol{A - Keys\ combined\ list} = \{\mathbf{20743}, \dots (\mathbf{48}\ \boldsymbol{terms}) \dots, \mathbf{12282}\} \quad (52)$$

$$\boldsymbol{B - Keys\ combined\ list} = \{\mathbf{20743}, \dots (\mathbf{48}\ \boldsymbol{terms}) \dots, \mathbf{12282}\} \quad (53)$$

$$K\ session\ key\ from\ Alice = \text{"}\boldsymbol{1f18d824829ebc368c13} \dots (total\ length\!: 64\ bytes)\text{"} \quad (54)$$

$$K\ session\ key\ from\ Bob = \text{"}\boldsymbol{1f18d824829ebc368c13} \dots (total\ length\!: 64\ bytes)\text{"} \quad (55)$$

### *4.4 The security of the RDMPF protocol depends on the following considerations:*

Generation of Private Matrices {X, Y. U, V} from public setup parameters (BaseXU and BaseYV). Because the goal is to achieve commutativity of these matrices, the mechanism must provide classical and quantum security. Note, however, that these matrices interpenetrate as exponents and in pairs (x, y); therefore, their impact on integral security is limited. One option is that the definition of the base matrices should consider the use of primitive characteristic polynomials of the matrices (in a semi-random selection) to ensure a long multiplicative order, thereby making it possible to define sufficiently long periods to avoid repetition over the power cycles.

The complexity of the RDMPF function can be approximated as if n=matrix dimension (and usually its rank is less than or equal to n-1). Each element of the resulting matrix uses n products of modular exponentiations of a known base, and exponents resulting from the inner product of two unknown vectors.

The naive bit complexity [20] of a product (k=ab) is O(log2 a. log2 b) and the modular exponentiation m=b^k (mod s) is O((log2 s)^2. log2 k) [24]. Therefore, assuming only unit complexity in the computation of each exponent, n-exponentiations span O(n (log2 s)^2. log2 k) bits and, using "square and multiple operations," the final computation of the resulting RDMPF matrix (proportional to n^2) will be O(n^3. (log2 s)^2 . log2 k) bits. Further complexity can be added using field operations (F2^p) instead of regular arithmetic [21]. The number of rounds does not improve the complexity but adds proportionally more time to any practical attack.

### *4.5 Parameters sensitivity of the RDMPK key-agreement in relative time units*

See Table 1. Some results obtained with unoptimized interpreted Mathematica code of the full RDMPF on an Intel Core i5-5200 U CPU @2.20 GHz, for different parameters and only one round (1 unit equals 265 ms).

**Table 1. Relative sensitivity computation of the RDMPF. The time values are the mean of 10 runs.**

| dimension | prime | expMax | relative time |
|---|---|---|---|
| Effect of a 5-fold dim change | | | |
| 5 | 997 | 1000 | 1 |
| 25 | 997 | 1000 | 577 |
| Effect of an approximate 5-fold prime change | | | |
| 5 | 997 | 1000 | 1 |
| 5 | 4973 | 1000 | 1.41 |
| Effect of a 5-fold expMax change | | | |
| 5 | 997 | 1000 | 1 |
| 5 | 997 | 5000 | 1.01 |

This table shows RDMPF's high sensitivity to dimensional change, relative sensitivity to prime change, and insensitivity to maximum exponent change due to the "square-and-multiply" algorithm [3] used.

*4.6 Using truly random integer matrices*

It is important to emphasize that the development of MPF represents a significant advance over the purely algebraic approach to PQC protocols owing to its inherent simplicity and security. Although the focus has primarily been on the overall structure, subtle refinements have improved both the efficiency and robustness. One of the most notable limitations of the original Matrix Power Function is the exclusive consideration of circulant matrices or those derived from special algebraic groups in pursuit of commutativity, rather than truly random integer matrices. This limitation was overcome using the algorithms presented in this paper.

## 5. Discussion

*5.1. Implementation suggestions.*

It is imperative to adapt this protocol to a Key Encapsulation Mechanism (KEM) to ensure that its intrinsic security complies with the NIST post-quantum standardization proposals [16]. This requires the addition of a postquantum public-key cryptosystem [1]. Other issues such as constant time or uniform power consumption should be considered to achieve side-channel attack protection [22]. The benefit achieved is the adaptation of semantic security to the IND-CCA level [25, 26, 27].

The following procedure presents the proposed KEM based on the multiRDMPF key-agreement protocol (KAP). Note that it cryptographically conceals all otherwise public information except the setup parameters, thereby enhancing security.

1. **ALGORITHMS USED:**

   multiRDMPF (RD:Rank-Deficient) MPF - KAP

   HMAC(key, SHA3-512)

2. **CHAINED NONCES**: η0 (shared secret, root nonce) , ηm(subsequent random nonces), all nonces have 512-bits. (Note: η0 can be obtained by a previous multiRDMPF KAP session or pre-distributed by separate channel, the nonces can be renewed using Lamport scheme [2]).

   *(|| concatenation symbol,* ⊕ bitwise XOR*)*

3. **AUTHENTICATION TAGS**: authA, authB, public values, of 256-bits

4. BOB Starts: (Obs: Points 4., 5., 6. are repeated round by round and the Token TA, TB are used as A-Token combined list and B-Token combined list following the multiRDMPF. Each TA, TB is padded to 512-bits.

   Bob has η0

   He generates PriB, TB (multiRDMPF KAP).

   Compute and Send CloseB to ALICE:

   CloseB= HMAC (η0, authA || authB) ⊕ TB

5. **ALICE (ENCAPSULATION):**

Alice has η0

Receives CloseB

Compute TB= CloseB ⊕ HMAC (η0, authA || authB)

Generate PriA, TA, KeyA (multiRDMPF KAP)

Compute

CloseA= HMAC (η0, authA || authB ⊕ ηm) ⊕ TA

Generate a random key K (512-bit)

Encapsulate

cipher = HMAC (KeyA, [authA || authB] ⊕ ηm) ⊕ K

Send {Encap, CloseA, ηm } to BOB

6. **BOB (DECAPSULATION):**

Receives {Encap, CloseA, ηm } from ALICE.

Retrieves

TA= CloseA ⊕ HMAC (η0, [authA || authB] ⊕ ηm).

Compute KeyB (multiRDMPF KAP)

Decapsulate

K = Encap ⊕ HMAC (KeyB, [authA || authB] ⊕ ηm).

## Summary and conclusions

The two protocols use rectangular or rank-deficient matrices instead of the original MPF matrices. Both block algebraic linearization and Gröbner basis attacks [28-33]. In addition, the second method, using rank-deficient square matrices that allow conventional linear algebra over singular matrices, significantly enhances internal complexity and adds resilience to algebraic attacks. The main objective is to introduce quantum-safe protocols using new algorithms that require neither special hardware nor extended precision. Consequently, they can easily be implemented on commercially available computers.

To safeguard against algebraic attacks [28-33] and guarantee a well-defined numerical format, we incorporate p-modular operations in our protocols. We increase the entropy of key search spaces by replacing circulant matrices [4-9] or restricted algebraic groups to achieve commutativity with unstructured random integers. An important aspect that should not be overlooked is the security of the pseudorandom bit generators. Therefore, we strongly recommend using algorithms that demonstrate resilience to side channels and quantum attacks, as described in [34]. One idea that can be implemented is the creation of a subliminal channel using an alternatively secure method, in which the parties privately agree upon a linear combination of rows to create rank-deficient matrices.

**Supplementary materials:** Mathematica 12 notebooks with all functions and numerical examples used in our KAP variants can be distributed upon request to phecht@dc.uba.ar

**Conflicts of Interest:** The authors declare no conflicts of interest.

## References

[1] Bernstein D. et al. (2009), Post-Quantum Cryptography, Springer, https://doi.org/10.1007/978-3-540-88702-7

[2] Nielsen M.A..; Chuang, I. L. (2010), Quantum computation and quantum information. Cambridge University Press

[3] Menezes, A. J., van Oorschot, P. C., & Vanstone, S. A. (1996). Handbook of applied cryptography, CRC Press

[4] Sakalauskas E. (2018), Enhanced matrix power function for cryptographic primitive construction. Symmetry, 10, 43.

[5] Sakalauskas E. et al. (2008), and the Key Agreement Protocol (KAP) Based on Matrix Power Function. In Advanced Studies in Software and Knowledge Engineering, Information Science and Computing, pp. 92–96.

[6] Sakalauskas, E.; Luksys, K. (2012) Matrix power function and its application to block cipher s-box construction, Int. J. Inn. Comp. Inf. Contr., 8, 2655–2664.

[7] Sakalauskas E. and Mihalkovich A., (2017), Improved Asymmetric Cipher Based on Matrix Power Function Resistant to Linear Algebra Attack. Informatica, 28, 517–524.

[8] Sakalauskas E. and Mihalkovich A. (2018), MPF Problem over Modified Medial Semigroup Is NP-Complete. Symmetry 10, 571.

[9] Sakalauskas E. and Mihalkovich A. (2020), A New asymmetric cipher of non-commuting cryptography class based on enhanced MPF, IET Information Security, Volume 14, Issue 4, 410-418

[10] Galbraith, S., (2012), The KEM/DEM paradigm, Mathematics of Public-Key Cryptography. Cambridge University Press. pp. 471-478. ISBN 978-1-107-01392-6

[11] Shor P.W. (1999), polynomial time algorithms for prime factorization, and discrete logarithms on a quantum computer, SIAM Rev. 41, 303–332.

[12] Zalka, C. (1999). Grover's quantum searching algorithm is optimal, Physical Review A, 60(4), 2746.

[13] McDonald, I., & Sweet, M. (2015). Internet Printing Protocol (IPP) Over HTTPS Transport Binding and Uri Scheme (No. rfc7472).

[14] Rempola, M. H., Smith, A., Li, Y., & Du, L. (2024, June). Securing SDN Communication through Quantum Key Distribution. In 2024, The IEEE Transportation Electrification Conference and Expo (ITEC) (pp. 1-5). IEEE.

[15] Scolnik H.D. and Hecht J.P., (2023) Post-Quantum Key Agreement Protocol based on Non-Square Integer Matrices, https://arxiv.org/abs/2301.01586

[16] NIST PQC, (2017-2024) https://csrc.nist.gov/projects/post-quantum-cryptography (consulted 10-25-2024)

[17] Marmebro, A., & Stenbom, K. (2024). Investigation of Post-Quantum Cryptography (FIPS 203 and 204) Compared to Legacy Cryptosystems, and Implementation in Large Corporations.

[18] Tan, W., Lao, Y., & Parhi, K. K. (2024, May). Area-efficient matrix-vector polynomial multiplication architecture for ML-KEM using interleaving and folding transformations. In 2024, The IEEE International Symposium on Circuits And Systems (ISCAS) (pp. 1-5). IEEE

[19] Rajendran, G., Ravi, P., D'anvers, J. P., Bhasin, S., & Chattopadhyay, A. (2023). Pushing the limits of generic side-channel attacks on LWE-based KEM-parallel PC oracle attacks on Kyber KEM and beyond, IACR Transactions on Cryptographic Hardware And Embed Systems

[20] Bach E., Shallit, J. (1997), Algorithmic Number Theory, Vol 1-Efficient Algorithms, MIT press

[21] Hecht, P. (2020), PQC: R-Propping of Public-Key Cryptosystems Using Polynomials over Non-commutative Algebraic Extension Rings, https://eprint.iacr.org/2020/1102, 10pp, DOI: 10.13140/RG.2.2.25826.56002

[22] Guo, Q., Nabokov, D., Nilsson, A., & Johansson, T. (2023, December). SCA-LDPC: A code-based framework for key-recovery side-channel attacks on post-quantum encryption schemes. At The International Conference on The Theory And Application of Cryptology and Information Security (pp. 203-236). Singapore: Springer Nature Singapore.

[23] Cramer, R.Shoup, V.( The design and analysis of practical public-key encryption schemes are secure against adaptive chosen ciphertext attacks. SIAM Journal of Computing. 33

[24] Jiang, H. et al, (2017) IND-CCA-secure Key Encapsulation Mechanism in the Quantum Random Oracle Model, Revisited, https://eprint.iacr.org/2017/1096.pdf

[25] Cramer, R., Shoup, V. (April 2002). Universal hash proofs and a paradigm for adaptively chosen ciphertext secure public key encryption. At The International Conference On The Theory And Applications Of Cryptographic Techniques (pp. 45-64). Berlin, Heidelberg: Springer Berlin Heidelberg.

[26] Barthe, G., Grégoire, B., Lakhnech, Y., & Zanella Béguelin, S. (2011, February). Beyond provable security: verifiable IND-CCA security of OAEP, Cryptographers' Track At The RSA Conference (pp. 180-196). Berlin, Heidelberg: Springer

[27] Jiang, H. et al, (2017) IND-CCA-secure Key Encapsulation Mechanism in the Quantum Random Oracle Model, Revisited, https://eprint.iacr.org/2017/1096.pdf

[28] Liu J. et al. (2014). A linear algebraic attack on a non-commuting cryptography class is based on the matrix power function. In International Conference on Information Security and Cryptology; Springer: Cham, Switzerland, 2016; pp. 343–354. matrix power function. Informatica, 25, 283–298.

[29] Roman'kov V., (2017), Cryptanalysis of a combinatorial public key cryptosystem. Groups Complexity Cryptology, 9(2), 125-135.

[30] Roman'kov V. (2016), Nonlinear decomposition attack. Groups Complexity Cryptology, 8(2), 197-207

[31] Myasnikov A., and V. V. Roman. (2015), A linear-decomposition attack. Groups Complexity Cryptology, 7(1), 81-94

[32] Myasnikov A. et al. (2011), Non-commutative Cryptography and Complexity of Group-theoretic Problems, Mathematical Surveys and Monographs, AMS Volume 177

[33] Ben-Zvi A. et al. (2018) and cryptanalysis using algebraic spans, Advances in Cryptology–CRYPTO 2018: 38th Annual International Cryptology Conference, Santa Barbara, CA, USA, August 19–23, Proceedings, Part I 38 (pp. 255-274). Springer International Publishing

[34] Di Mauro, J., Salazar, E., and Scolnik, H. D. Design and implementation of a novel cryptographically secure pseudorandom number generator. J Cryptogr Eng 12, 255–265 (2022). https://doi.org/10.1007/s13389-022-00297-8